\documentclass[twoside]{dis07}
\usepackage[latin1]{inputenc}
\usepackage[dvips]{graphicx,epsfig,color}
\usepackage{wrapfig,rotating}
\usepackage{amssymb,amsmath,array}

\begin{document}
\title{Charm at CLEO-c}

\author{Kamal K. Seth\\
(for the CLEO Collaboration)
\vspace{.3cm}\\
%
Northwestern University - Department of Physics \& Astronomy \\
Evanston, IL 60208 - USA
}

\maketitle

\begin{abstract}
A minireview of the recent results from CLEO-c is presented.  It includes new results in charmonium spectroscopy, charmonium-like exotics, and open-charm decays.
\end{abstract}

\sloppy

\section{Introduction}

During the last 25 years, CLEO was primarily devoted to the physics of bottomonium ($b\bar{b}$) and $B$--mesons.  CLEO and CESR have recently morphed into CLEO-c and CESR-c to do research in the charm quark region, $\sqrt{s}=3-5$~GeV to address challenging questions in charmonium spectroscopy, meson form factors, and spectroscopy of $D$ and $D_s$ mesons.

\section{Spin--singlet states of Charmonium ($c\bar{c}$)}

The spin--triplet states of charmonium ($^3S_1(J/\psi,\psi')$, $^3P_J(\chi_{c0,c1,c2})$) have been extensively studied by SLAC, Fermilab, and BES during the last 30 years.  The spin--singlet states have largely defied identification and study.  Neither $\eta_c'2^1S_0)$, nor $h_c(1^1P_1)$, nor $\eta_b(1^1S_0)$ have ever been convincingly identified.  This leaves us largely in the dark about the character of the all--important spin--spin hyperfine interaction between two quarks.  This serious shortcoming has now been mended by the successful identification of both $\eta_c'$ and $h_c$ at CLEO.

\subsection{Identification of the $\eta_c'(2^1S_0)$ state of Charmonium}

The hyperfine splitting of the $1S$ state of charmonium is known to be $\Delta M_{hf}(1S)\equiv M(J/\psi)-M(\eta_c)=116.5\pm1.2$~MeV.  While the mass of $\psi'(2^3S_1)$ is well known, the lack of any knowledge of $\eta_c'(2^1S_0)$ has prevented us from knowing how the spin--spin interaction changes for the radially excited states.  A study by Belle~\cite{belle-etacp} of the decays of 45 million $B$ mesons, $B\to K(K_SK\pi)$, gave the first hint that the $\eta_c'$ mass was substantially larger than expected. In a recent measurement, CLEO has confirmed this and has successfully identified $\eta_c'$ in its formation in two photon fusion, and its decay into $K_SK\pi$ \cite{cleoc-etacp}. Its mass spectrum, shown in Fig.~1~(left), leads to $\Delta M_{hf}(2S)=43.1\pm3.4$~MeV, which is almost a factor three smaller than $\Delta M_{hf}(1S)$, and its explanation remains a challenge for the theorists.


\begin{figure}
\begin{center}
\includegraphics[width=2.8in]{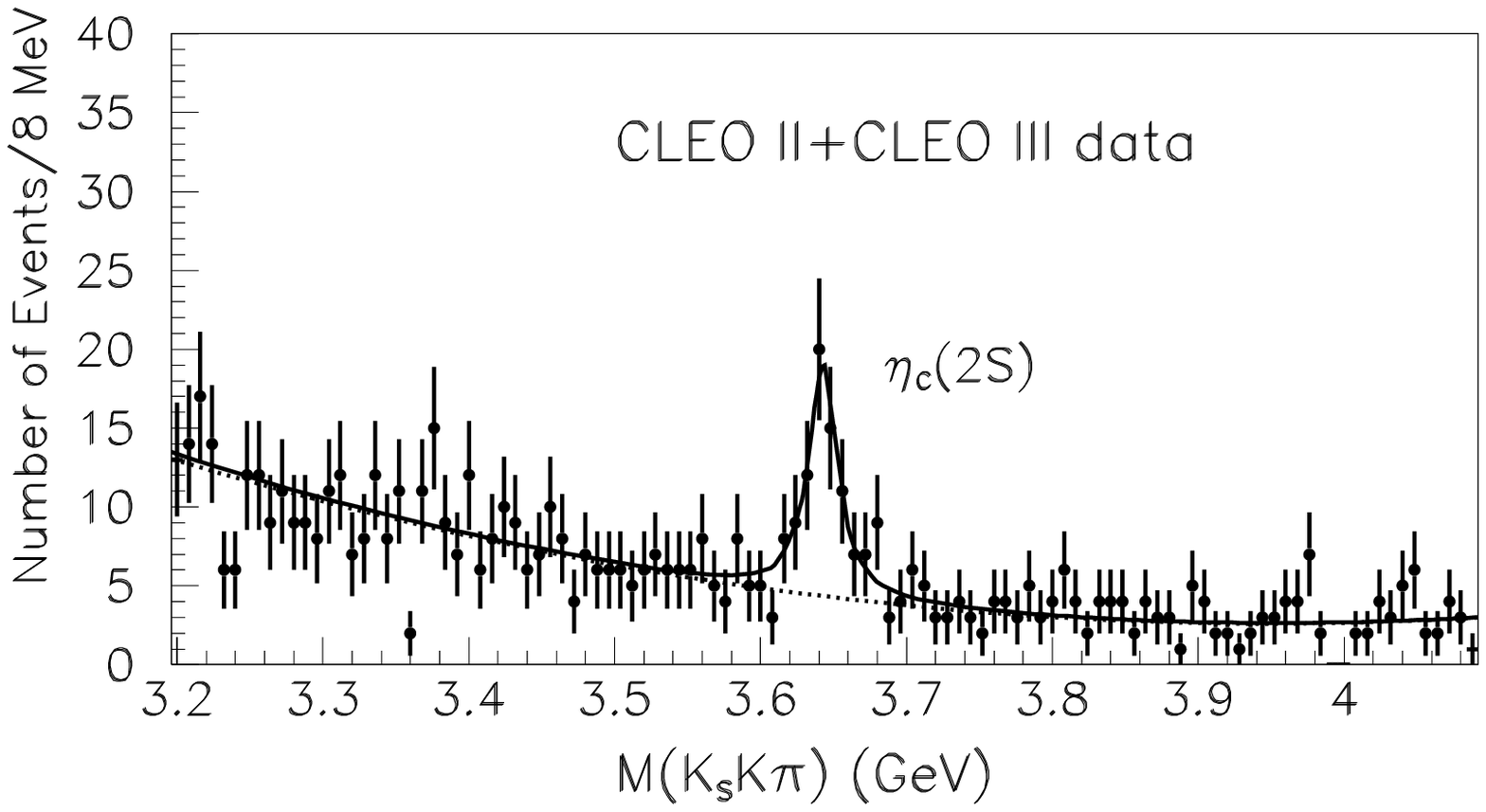}
\raisebox{1.5in}{\rotatebox{270}{\includegraphics[width=1.4in]{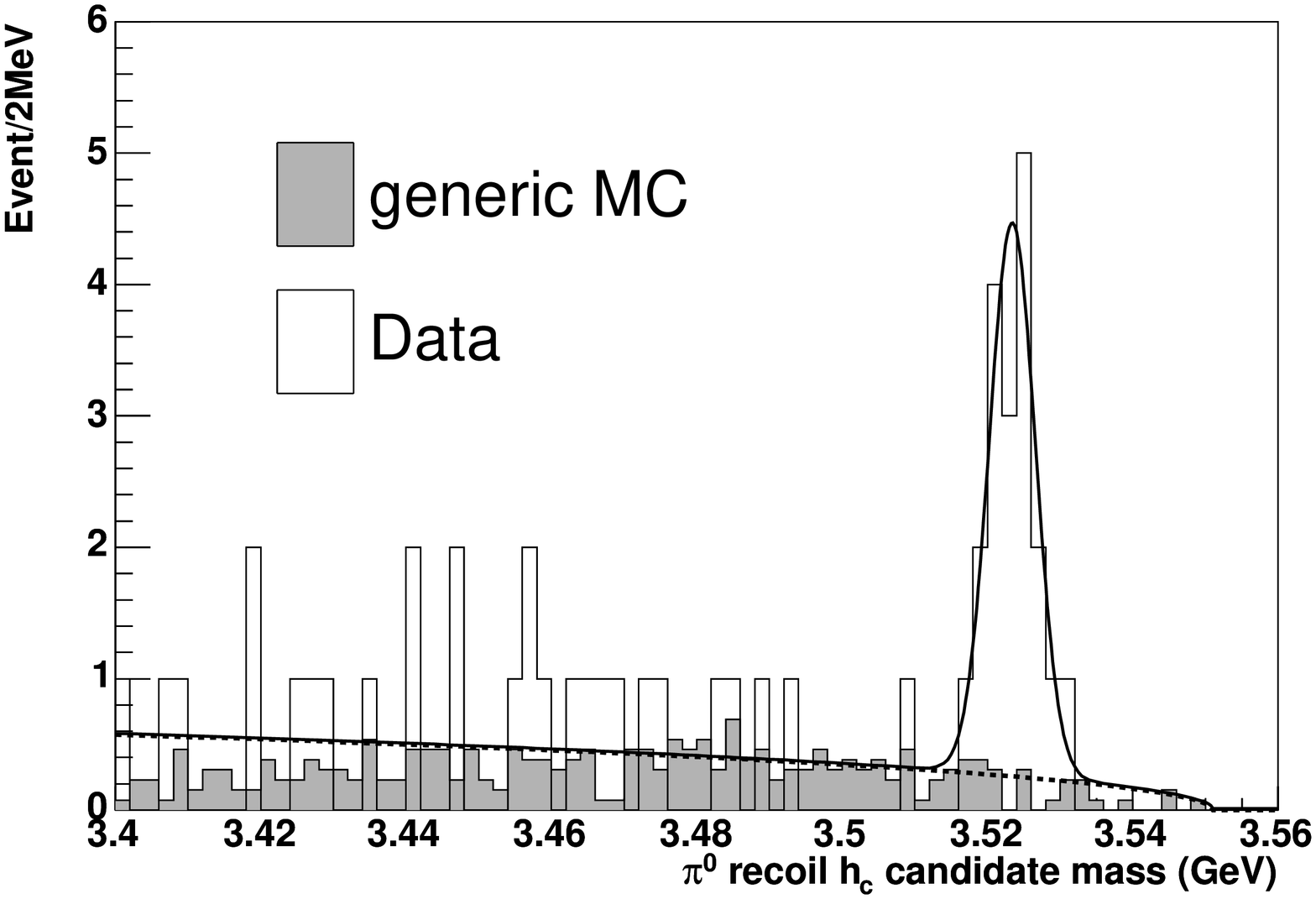}}}
\end{center}
\caption{(left) $M(K_SK\pi)$ from CLEO showing $\eta_c(2S)$.  (right) $\pi^0$ recoil mass spectrum from CLEO showing $h_c$ in exclusive analysis of $\psi(2S)\to\pi^0h_c,~h_c\to\gamma\eta_c$. }
\end{figure}

\subsection{Identification of the $h_c(1^1P_1)$ state of Charmonium}

It is of great interest to determine how the hyperfine interaction changes with the orbital angular momentum between two quarks.  In the generally accepted potential models, there is no long--range spin--spin component and the consequent prediction is that $\Delta M_{hf}=0$ for all $l\ne0$.  In order to test this prediction for $l=1$, one has to identify $h_c$, the singlet $P$ state, since the triplet $P$ states, $\chi_{cJ}$ are well known.  Unfortunately, the radiative transition $\psi'\to\gamma h_c$ is forbidden by $C$--parity, and $h_c$ has defied firm identification despite numerous earlier attempts. In a very challenging measurement of the isospin forbidden reaction,\\ $\psi'\to\pi^0 h_c,$ $h_c\to\gamma\eta_c$, CLEO \cite{cleoc-hc} has recently unambiguously identified $h_c$ in both inclusive and exclusive analyses of their data for 3 million $\psi'$.  The mass spectrum for the exclusive reaction is shown in Fig.~1~(right), where the $h_c$ peak  is clearly seen.  The result is $\Delta M_{hf}(1P)=+1.0\pm0.6\pm0.4$~MeV.  CLEO-c has now taken data with 24.5 million $\psi'$, and a $h_c$ peak with $\sim250$ counts is expected, which will reduce both errors by more than a factor of two.

The  determination of the $^3P_J$ centroid as $M(\left<^3P_J\right>)=M(5\chi_{c2}+3\chi_{c1}+\chi_{c0})$ appears to lead to $\Delta M_{hf}(1P)\approx0$, but J.-M.~Richard \cite{richard} has pointed out that a more ``correct'' determination of the centroid implies $\Delta M_{hf}(1P)\approx4$~MeV.

\section{Hadronic decays of $\chi_{cJ}$ states}

Recently, extensive analyses of CLEO data with 3 million $\psi'$ has been done for $\chi_{cJ}$ 2, 3, 4--body decays \cite{cleo-chic}. Analysis of our  new sample of 24.5 million $\psi'$ will greatly improve the precision of these results and increase knowledge of $\chi_{cJ}$ decays several-fold.

\section{Timelike form factors of pion and kaon}

Using 21 pb$^{-1}$ of $e^+e^-$ annihilation data taken off of the $\psi'$ resonance, CLEO-c has made the world's first precision measurements of the timelike form factors of charged pions and kaons at $|Q|^2=13.48~\mathrm{GeV}^2$ \cite{cleo-ff}.  The measurements show that there is essentially no theoretical understanding of timelike form factors of mesons at present.

\section{Charmonium--like states}

As is well known, there has been a ``renaissance'' in hadron spectroscopy during the last couple of years with reports of one unexpected resonance after another by Belle and BaBar who are able to play in the game with hundreds of inverse femtobarns of luminosity.  It began with X(3872), then X, Y, Z(3940), and then Y(4260).  CLEO has nothing to say about X, Y, Z(3940), but it has made contributions to the study of X(3872) and Y(4260).

\subsection{X(3872)}

Belle discovered it, and CDF and D\O~have confirmed it.  The state has $M$(X(3872))$=3871.2\pm0.5$~MeV, and a very narrow width, $\Gamma($X(3872))$<2.3$~MeV.  Numerous theoretical conjectures about the nature of this state, which prominently decays into $\pi^+\pi^-J/\psi$, have been made.  The most popular among these is that it is a $D^0\overline{D^{*0}}$ molecule, inspired by the fact that its mass is very close to $M(D^0)+M(\overline{D^{*0}})$.  It occured to us that for this model to survive, it is very important to know what the molecule's binding energy is. This requires an accurate measurement of the mass of $D^0$.  At CLEO-c, we have made a precision measurement of $M(D^0)$ \cite{cleo-d0mass} by means of the reaction $e^+e^-\to\psi(3770)\to D^0\overline{D^0}$, $D^0\to K_S\phi$, with the result that $M(D^0)=1864.847\pm0.178$~MeV, and hence $E_B($X(3872)$=+0.6\pm0.6$~MeV.  While this small binding energy allows X(3872) to be bound, it results in a prediction of its width for decay into $D^0\overline{D^0}\pi^0$ which is a factor $\sim200$ smaller than that observed by Belle \cite{belle-xddpi0}.  This could be a death--blow to the molecular model.

\begin{figure}
\begin{center}
\includegraphics[width=2.2in]{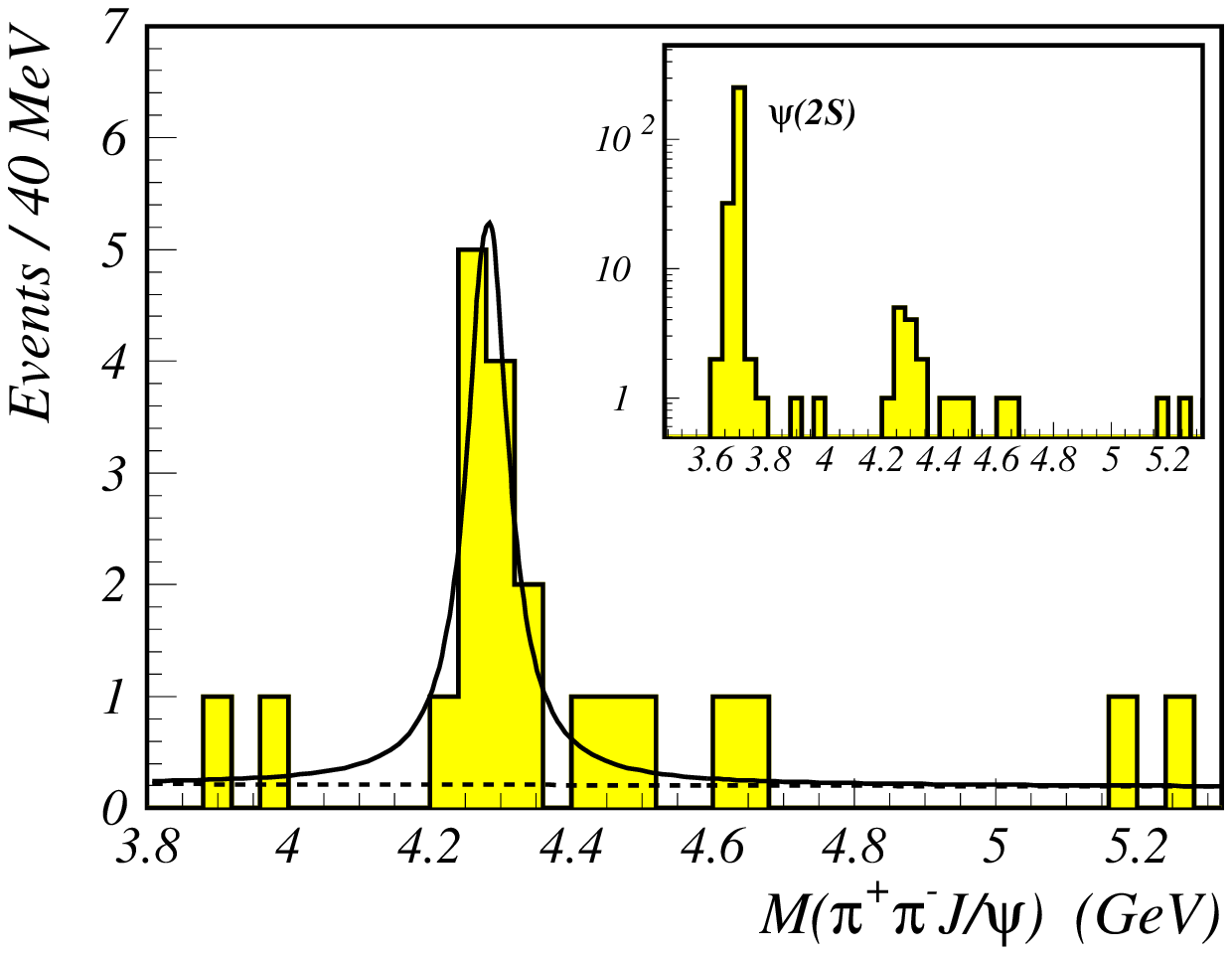}
\includegraphics[width=1.5in]{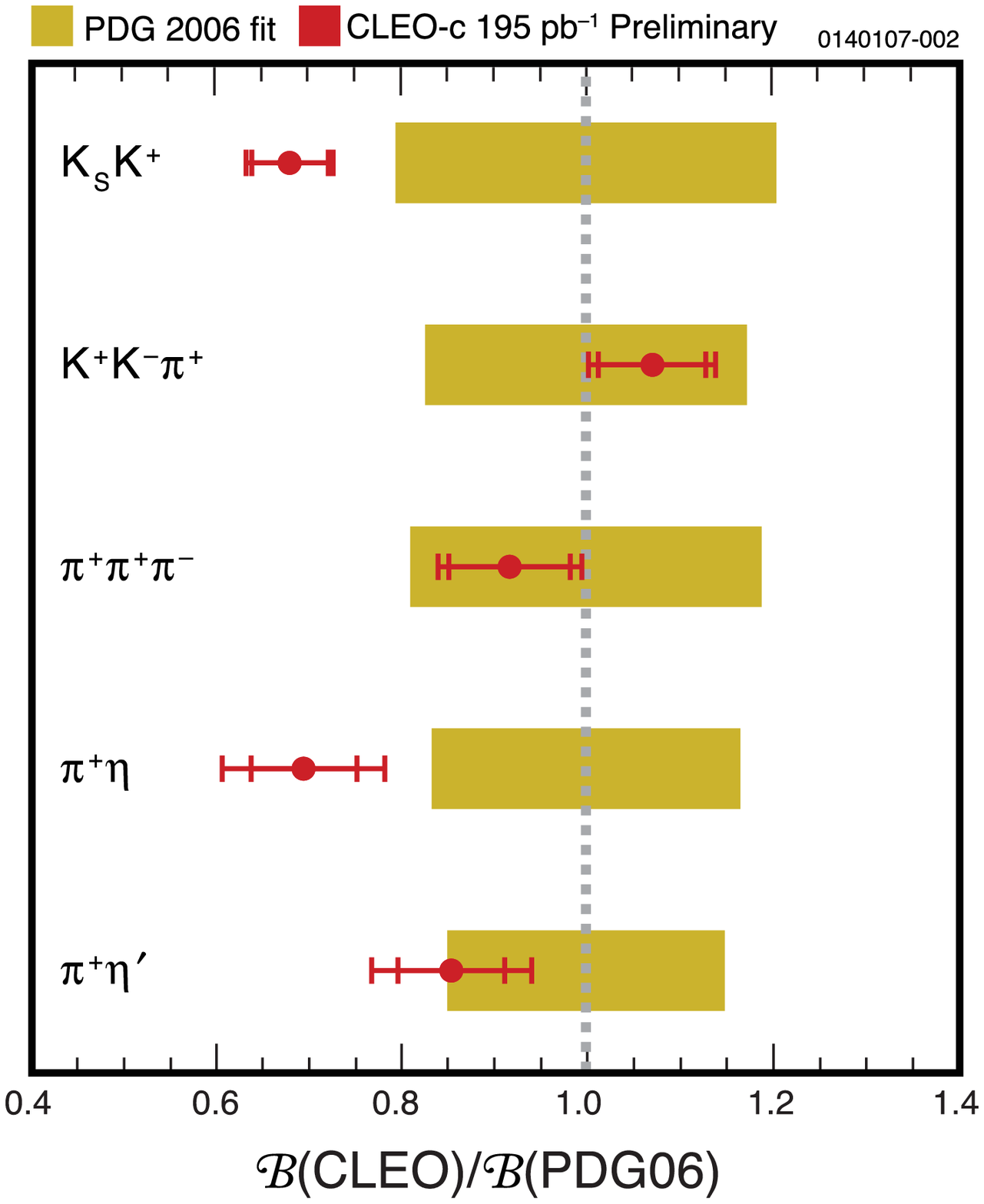}
\includegraphics[width=1.65in]{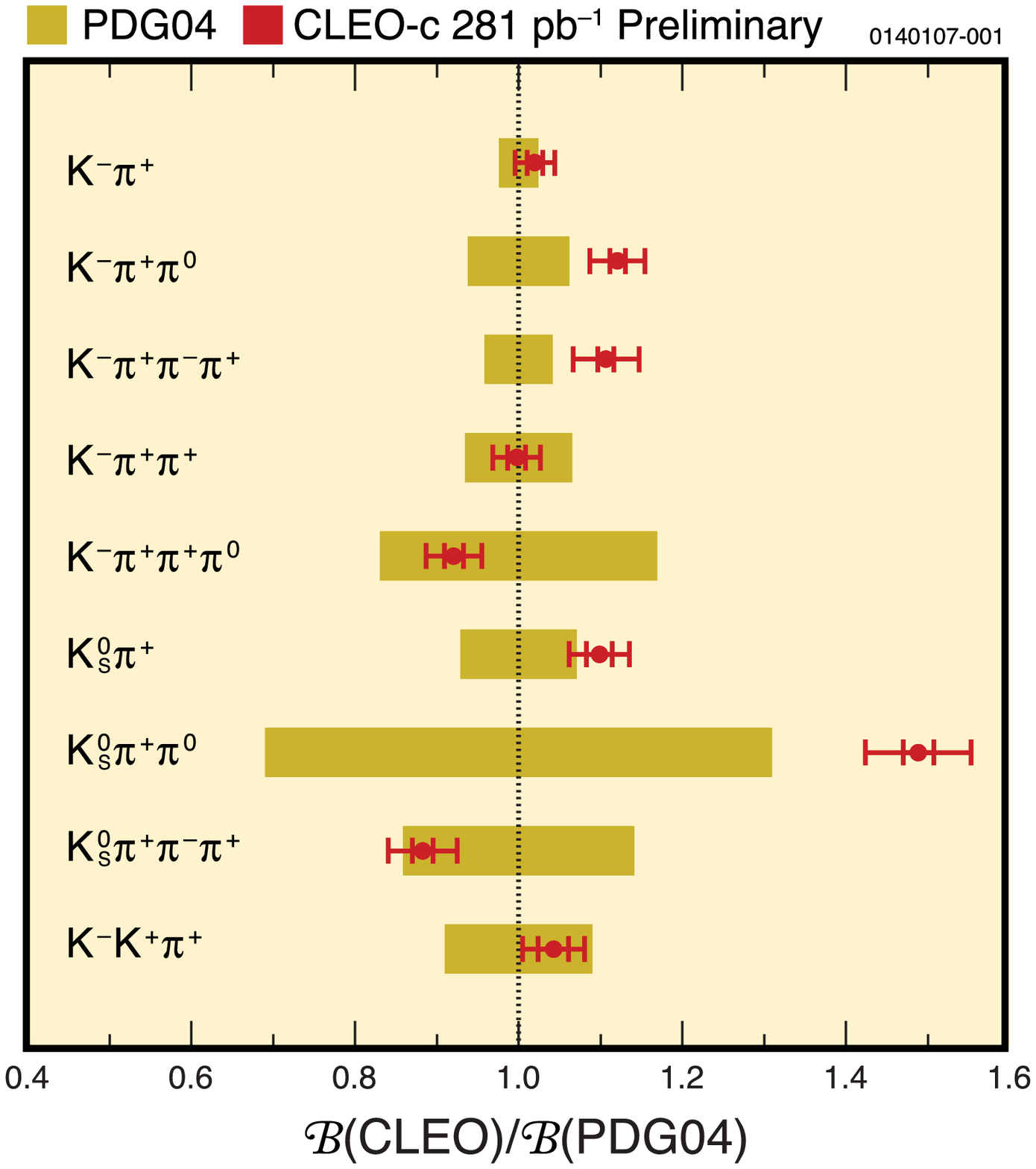}
\end{center}
\caption{(left) CLEO observation of Y(4260) in two photon fusion.  (middle \& right) Ratios of CLEO-c results for hadronic decays of $D$ and $D_s$ (dots without bars) to the current PDG values (with shaded error bars).}
\end{figure}

\subsection{Y(4260)}

BaBar \cite{babar-y} has reported observing a resonance Y(4260) with $M($Y$)=4259\pm8^{+2}_{-6}$~MeV,~$\Gamma($Y$)=88\pm23^{+6}_{-4}$~MeV, in ISR production, $e^+e^-\to\gamma_{ISR} e^+e^-\to\gamma_{ISR}(\pi^+\pi^-J/\psi)$.  At CLEO, despite a factor $\sim20$ smaller luminosity, we observe (see Fig.~2(left)) a clear signal for Y(4260) with very small background \cite{cleoc-y}.  The ISR observation of the resonance confirms its vector nature.  This is rather bizarre because the $R\equiv\sigma(h\bar{h})/\sigma(\mu\mu)$ measurements show a deep minimum at $\sqrt{s}=4260$~MeV, instead of a maximum expected for a vector state.  The resonance is therefore rather mysterious.

\section{CLEO-c as an open--charm factory}

The primary motivation for CLEO to morph into CLEO-c was to become a prodigious factory for the production of open--charm hadrons, the $D$ and $D_s$, and thereby enable it to make important contributions to $D$ physics, to determine form factors, CKM matrix elements, and to allow peeks into the holy--grail of \textit{``beyond the standard model.''}

CLEO-c has started this program very successfully by taking $\sim540$~pb$^{-1}$ of data at $\psi(3770)$ with near threshold production of $D\overline{D}$, and $\sim313$~pb$^{-1}$ at $\psi(4170)$ for near-threshold production of $D_s\overline{D_s}$.  A large number of precision measurements of hadronic decays of $D$ and $D_s$ have already been made \cite{cleo-dhad}.  These are illustrated in Fig.~2(right).

Leptonic decays of $D$ and $D_s$  have been measured to obtain $\Gamma(D^+_{(d,s)}\to l^+\nu)$.  Using the best known values of the CKM matrix elements, the decay constants $f(D^+)$and $f(D_s^+)$ have been deduced \cite{cleo-lep}.  The results, listed in Table I, agree very well with the latest unquenched lattice predictions.

Semi-leptonic decays of $D$ mesons $D^0\to(\pi^-,K^-)e^+\nu$ and $D^+\to(\pi^0\overline{K^0})e^+\nu$ have also been successfully measured, and using the form factors predicted by unquenched lattice calculations, the CKM matrix elements $|V_{cd}|$ and $|V_{cs}|$ have been obtained in agreement with their unitarity--based values.

\begin{table}
\begin{center}
\begin{tabular}{r|c|c}
\hline \hline
 & CLEO-c & Unquenched Lattice \\
\hline
$f(D^+)$ & $222.6\pm16.7^{+2.8}_{-3.4}$ MeV & $201\pm3\pm17$ MeV\\
$f(D_s^+)$ & $280.1\pm11.6\pm6.0$ MeV & $249\pm3\pm16$ MeV\\
$f(D^+_s)/f(D^+)$ & $1.26\pm0.11\pm0.03$ & $1.24\pm0.01\pm0.07$\\
\hline \hline
\end{tabular}
\end{center}
\caption{Measured $D$ and $D_s$ decay constants compared to unquenched lattice predictions.}
\end{table}

\section{Summary}

To summarize, CLEO has made a transition to CLEO-c and is very successfully contributing to the study of the hidden flavor physics of the charmonium region and the open flavor physics of $D$--mesons.

\begin{footnotesize}



%

\end{footnotesize}


\end{document}